\documentclass[aps,prl,twocolumn,showpacs,showkeys,superscriptaddress]{revtex4-1}

\usepackage {amssymb}
\usepackage {amsmath}
\usepackage {graphicx}
\usepackage {longtable}
\usepackage {color}

\begin{document}
\title{Antiferromagnetic resonance in ferroborate NdFe$_3$(BO$_3$)$_4$}

\author{M.I.Kobets}
\affiliation{B. Verkin Institute for Low Temperature Physics and
Engineering, National Academy of Sciences, Kharkov 61103, Ukraine}

\author{K.G. Dergachev}
\affiliation{B. Verkin Institute for Low Temperature Physics and
Engineering, National Academy of Sciences, Kharkov 61103, Ukraine}

\author{E.N. Khatsko}
\email[]{khatsko@ilt.kharkov.ua}
\affiliation{B. Verkin Institute for Low Temperature Physics and
Engineering, National Academy of Sciences, Kharkov 61103, Ukraine}

\author{S.L. Gnatchenko}
\affiliation{B. Verkin Institute for Low Temperature Physics and
Engineering, National Academy of Sciences, Kharkov 61103, Ukraine}

\author{L.N. Bezmaternykh}
\affiliation{L. Kirensky Institute of Physics of SB RAS, 660036 Krasnoyarsk, Russia}

\author{V.L. Temerov}
\affiliation{L. Kirensky Institute of Physics of SB RAS, 660036 Krasnoyarsk, Russia}

\begin{abstract}
The AFMR spectra of the NdFe$_3$(BO$_3$)$_4$ crystal are measured in a
wide range of frequencies and temperatures. It is found that by the type
of  magnetic anisotropy the compound is an "easy-plane" antiferromagnet
with a weak anisotropy in the basal plane. The effective magnetic parameters
are determined: anisotropy fields $H_{a1}$=1.14 kOe and $H_{a2}$=60 kOe and
magnetic excitation gaps $\Delta\nu_1$=101.9 GHz and $\Delta \nu_2$=23.8 GHz.
It is shown that commensurate-incommensurate phase transition causes a shift
in resonance field and a considerable change in absorption line width.

At temperatures below 4.2 K nonlinear regimes of AFMR excitation at low
microwave power levels are observed.
\end{abstract}  
 \maketitle

\section{Introduction}
The class of materials with coexistent magnetic and electric ordering is
currently under deep investigation  \cite{lib1}. Among such compounds is a family of
ferroborates (RFe$^{3+}$(BO$_3$)$_4$); (R being a rare-earth ion). All representatives
of this family are multiferroics. The diversity of magnetic properties of
ferroborates is caused by the existence of two magnetic subsystems: Fe ions
and rare-earth ions. The specificity of interaction between rare-earth and
iron ions dictates the orientation of Fe magnetic moments in the ordered
state with respect to crystallographic axes. Of considerable importance on
such systems are magnetoelastic and magnetoelectric interactions. Moreover,
because of their nonlinear properties, such compounds are of great interest
and show promise for laser technique. Therefore, much research is devoted to
structural, magnetic, magnetoelectrical and magnetoelastic properties of the
ferroborates family (e.g. see reviews  \cite{lib2,lib3}).

At high temperatures all crystals of the RFe$^{3+}$(BO$_3$)$_4$ family have a
trigonal structure of khanty mineral CaMg$_3$(CO$_3$)$_4$ that belongs to a
noncentro-symmetric space group {\bf R$32$} ({\bf D$^7_3$}) \cite{lib4}. The structure fragments
are shown Fig. 1. The edge-connected spiral chains of the FeO$_6$
octahedra extended along the \textbf{\textit{c}} axis  are the basic elements of the structure.

\begin{figure}[t]
\begin{center}
\includegraphics[width=70mm]{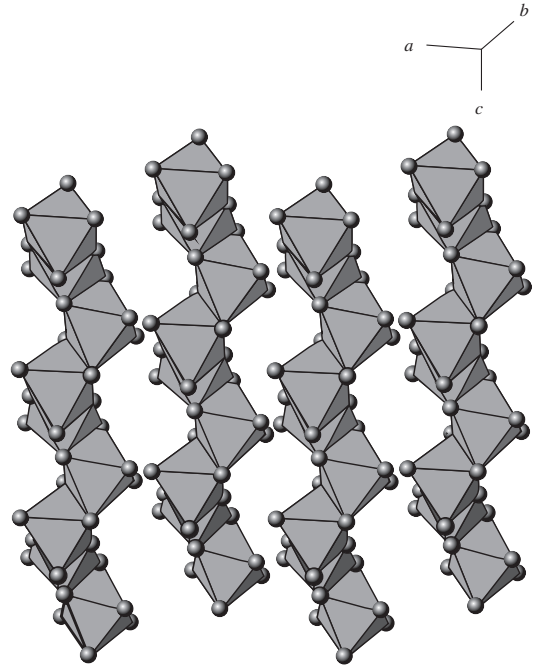}
\end{center}
\caption{Fragments of the NdFe$_3$(BO$_3$)$_4$ crystal structure.}
\end{figure}

The paper concerns the investigation of NdFe$_3$(BO$_3$)$_4$. The antiferromagnetic
ordering of the Fe subsystem in NdFe$_3$(BO$_3$)$_4$ occurs at 30 K and it is
followed by unit cell doubling along the \textit{\textbf{c}} axis  The magnetic cell of
ferroborate is inconsistent with the crystallographic one \cite{lib5}. The situation with Nd
subsystem is not so clear. Some authors believe \cite{lib6,lib7}that both subsystems are ordered at the same temperature, some authors \cite{lib8} claim that subsystem of Nd is just magnetizing by Fe

The intra - and interchain interactions in the Fe subsystem are antiferromagnetic
and equal to 580 and 270~kOe, respectively \cite{lib5}. Detailed study
of magnetization along \textbf{\textit{a}}, \textbf{\textit{b}} and \textbf{\textit{c}} axes  at fields up to 100 kOe
and T=2 K are reported in Ref. \cite{lib8}. At is shown that for the trigonal
antiferromagnet NdFe$_3$(BO$_3$)$_4$ the anisotropy of Fe and Nd subsystems
causes the magnetic moments of Fe$^{3+}$ and Nd$^{3+}$ to be oriented in parallel
in the basal plane, i.e. NdFe$_3$(BO$_3$)$_4$ is a magnet with an "easy-plane"
magnetization anisotropy. The existence of the triad axis gives rise to three
types of domains in zero field with the axes of antiferromagnetism at an angle
120$^\circ$. The antiferromagnetism vectors {\bf l} are oriented along binary
axes {\bf C$_2$}. On one of the domains the antiferromagnetism vector direction
coincides with the crystallographic axis {\bf a}. At T=4.2 K and H$_{sf}$=10 kOe
there occurs a spin-flop transition to a state where magnetic moments are
almost perpendicular to magnetic field. 

The paper \cite{lib9} reports that a jump-like initiation of magnetostriction
and electrical polarization is observed at magnetic fields corresponding to the
spin-flop transition at is found that in this case the crystal symmetry is
changed from class {\bf R$32$} to class {\bf R$2$} (monoclinic system).

The recent paper dealing with neutron scattering in NdFe$_3$(BO$_3$)$_4$ reports
that at T=13.5~K one can observe a magnetic phase transition from commensurate
orderedmagnetic phase to incommensurate long periodic helicoidal magnetic
superstructure\cite{lib7}.

Despite considerable interest of experimenters to ferroborates, the resonance
properties of the family remain almost unknown. There are only two reports on
antiferromagnetic resonance (AFMR) in GdFe$_3$(BO$_3$)$_4$ and YFe$_3$(BO$_3$)$_4$
\cite{lib11,lib12}.

The main concern of our paper is investigation of resonance properties of
NdFe$_3$(BO$_3$)$_4$ in paramagnetic and magnetoordered states by using
radio-frequency spectroscopic methods (EPR and AFMR).

\section{Experiment results and discussion}
The measurements of magnetic resonance in paramagnetic and ordered states
(T$_N$=30~K) of the crystal for three principal lattice directions were conducted
at liquid helium temperature and above with the use of the equipment described
in \cite{lib11}.

The EPR spectrum were taken at frequencies of 21.7 and 76 GHz at T=60--100 K
and external field {\bf H} oriented along \textbf{\textit{a}} and \textbf{\textit{c}} axes. Under such conditions
one can observe a single broad EPR line of Fe$^{3+}$ ion with a g-factor
equal to 2. The spectrum is shown Fig. 2. The EPR spectrum of ion
Nd$^{3+}$, the g-factor of which is more than 2, is not observed for all permanent
field directions because it seems likely to be very broadened due to spin-orbital
coupling. The spectrum is observed only at low temperatures (20--30 K) but in this
temperature range the material is already magnetically ordered. Moreover, as given in
\cite{lib11,lib14}, the ground doublet of multiplet {$^4$\bf I$_{9/2}$} of Nd$^{3+}$
ions is exchange splitted by 8.8 cm$^{-1}$. This suggests that the EPR spectrum of
Nd is not observed in the frequency range 17--142 GHz and at T=4.2 K and higher.

\begin{figure}[t]
\begin{center}
\includegraphics[width=90mm]{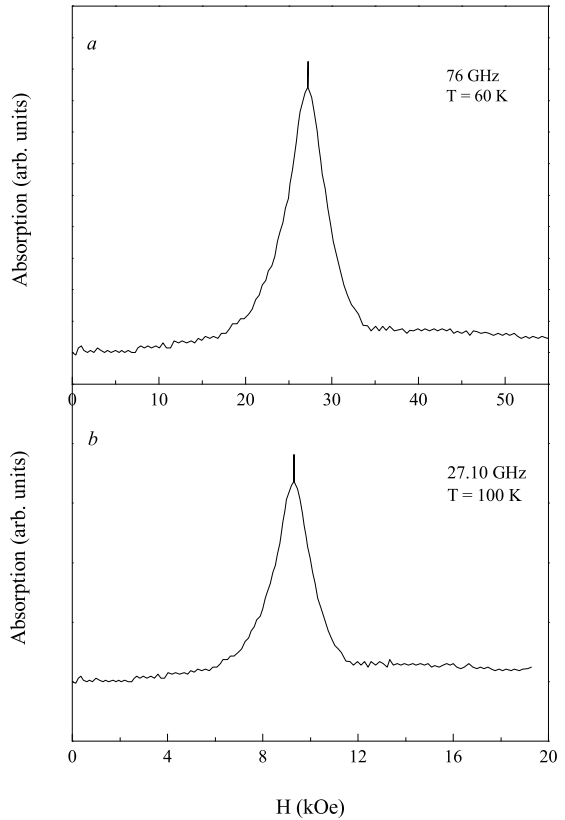}
\end{center}
\caption{The EPR spectrum of NdFe$_3$(BO$_3$)$_4$.
   a) $\nu$=76 GHz, T=60 K; b) $\nu$=27.1 GHz, T=100 K.
The narrow line at the curve apex is a DPPH signal.}
\end{figure}

As mentioned above in the papers cited, the ordered compound NdFe$_3$(BO$_3$)$_4$
is an "easy-plane" antiferromagnet. On the framework of two-sublattice magnet a
typical frequency field dependence of the AFMR spectrum for "easy-plane" is
schematically shown Fig. 3. For {$\bf H\parallel~\bf l\perp\bf z$} the AFMR
spectrum of such magnets consists of two considerably different branches one of
which is a low-frequency quasi-ferromagnetic gapless branch ($\nu_3$) and
the other is a high-frequency branch with energy gap $\sqrt{2H_eH_a}$ ($\nu_3$).
This branch becomes zero in magnetic field 2H$_e$. Besides, these branches have a point of
intersection (degeneration). The magnetic field in which the two branches are
crossed is arbitrarily called in literature as a field of spin-flop transition.
The field value is $\sqrt{2H_eH_a}$ (the value of the field which there occurs
an intersection of AFMR branches and the gap value are almost equal). When
external magnetic field is directed along the principle anisotropy axis
{$\bf H\parallel\bf l\perp\bf z$}, one of the AFMR branches is a rising quadratic
branch ($\nu_1$) and the other is not active.

\begin{figure}[t]
\begin{center}
\includegraphics[width=90mm]{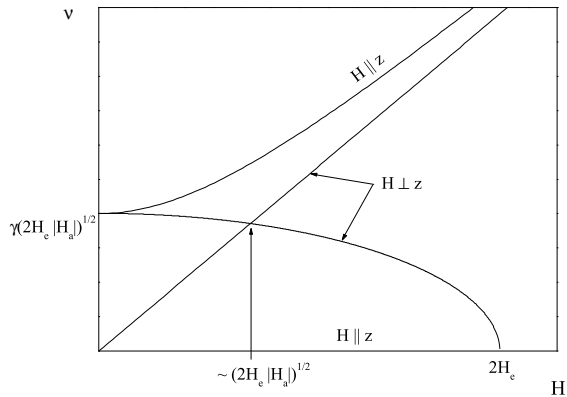}
\end{center}
\caption{Resonance frequency as a function of magnetic field for an
antiferromagnet with easy plane magnetic anisotropy.\cite{lib15}}
\end{figure}

The AFMR absorption spectra were measured in the following sequence. Since at
T=13.5~K there occurs a magnetic phase transition to a noncommensurate phase,
all frequency-field dependences of the AFMR spectrum were taken above and
below this transition at two temperatures (T=4.2 and 14.6 K).

At T=4.2 K we made a search for a high-frequency branch of the AFMR spectrum.
On the case of trigonal symmetry the magnetic axes {\bf x}, {\bf z} coincide with the
crystallographic axes \textbf{\textit{a}},\textbf{\textit{c}}. For frequencies ranged from 17.0 to 102.0 GHz
and external magnetic field perpendicular to the basal plane and parallel to
the principal anisotropy axis {\bf C$_3$} ({$\bf H\parallel \bf c$}), the antiferromagnetic
excitation is not observed. The high-frequency AFMR branch $\nu_1$(H) was
observed at 101.9 GHz in practically zero magnetic field (see Fig. 4).
This frequency represents an initial splitting (or a gap of magnon excitations)
for this branch. It equals {$\Delta\nu_{high}=\sqrt 2H_eH_a$=101.9 GHz}.
With $H_e$=580 kOe \cite{lib5}, we obtain the anisotropy value $H_a$=1.14 kOe.
The AFMR high-frequency branch rises quadratically with magnetic field. Its
frequency-field dependence is shown in Fig. 4 for T=4.2 and 14.6 K,
i.e. below and above the point of transition to an incommensurate phase.On the
model of two-sublattice antiferromagnet this dependence is given by the
expression:

\begin{equation}
(\nu / \gamma)^2=2H_eH_a+H^2,\label{equation1}
\end{equation}

\begin{figure}[t]
\begin{center}
\includegraphics[width=90mm]{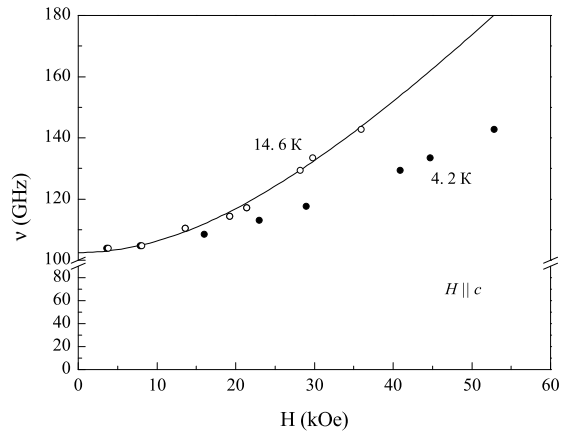}
\end{center}
\caption{The frequency-field dependences of the NdFe$_3$(BO$_3$)$_4$
AFMR spectrum for external magnetic field {$\bf H\parallel \textbf{\textit{c}}\perp\bf l$}
at two temperatures (4.2 and 14.6 K) below and above the transition to an
incommensurate phase. The solid line presents the data calculated by Eq. (1).}
\end{figure}

As seen from Fig. 4, the experimental field dependence at 14.6 K (i.e.
in the region of collinear structure existence) is well described by theoretical
relation (1), while at 4.2 K (in incommensurate phase) Eq. (1) does not describe
the experimental dependence {$\nu(H)$}, even though the gap value in both phases is
equal. This provides support for the statement in \cite{lib7} that below 13.5 K
the magnetic structure is no longer a two-sublattice collinear one.

For magnetic field orientation {$\bf H\parallel \textbf{\textit{c}}\perp\bf l$} only one oscillation
branch is observed and other AFMR excitations are not revealed.

The frequency-field dependences of the AFMR spectrum at the external magnetic
field oriented in the basal plane along \textbf{\textit{a}} axis ({$\bf H\parallel \textbf{\textit{a}}$}) are shown
in Fig. 5. As seen from the figure, there are two AFMR resonance lines
at this field orientation. The frequency of one of these lines {$\nu_2$} has an
initial value in zero magnetic field, {$\Delta\nu_2$=101.9 GHz}, determined by the
existence of a gap in the excitation spectrum. It is very difficult to observe
the AFMR spectrum of this branch at strict orientation ({$\bf H\parallel\bf l$})
because its frequency is weakly dependent on magnetic field and to excite
oscillations at the resonance frequency requires parallel orientation of constant
and high-frequency magnetic field ({$\bf H\parallel\bf h$}). But in our case it is
just observed and, as shown in Fig. 3, it corresponds to a high-frequency
$\nu_2$(H) branch (sometimes it is called in literature as exchange one).

\begin{figure}[t]
\begin{center}
\includegraphics[width=90mm]{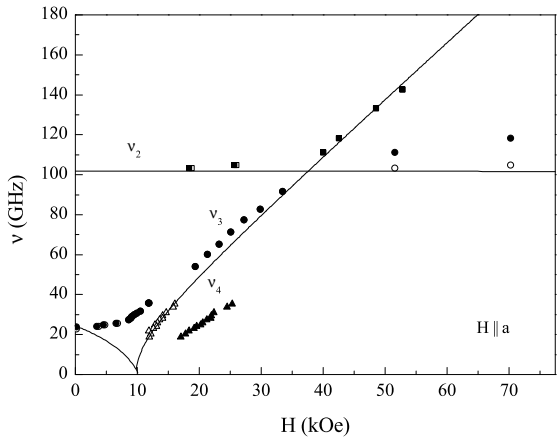}
\end{center}
\caption{The frequency-field dependences of AFMR spectrum for external
magnetic field {$\bf H\parallel \textbf{\textit{a}}\parallel\bf l$} at T=4.2 K
(black circles and squares) and T=14.6 K (open circles and squares).
The black and open triangles refer to resonance fields of spin-flop mode
at the same temperatures.}
\end{figure}

The second resonance line corresponds to a low-frequency gapless (quasi-ferromagnetic)
branch {$\nu_3$}(H) which is of nonlinear behavior in NdFe$_3$(BO$_3$)$_4$.

The above behavior of the frequency-field dependences of the AFMR spectrum
at ({$\bf H\parallel \textbf{\textit{a}}$}) is determined by the following conditions. As followed
from Fig. 3, when the external magnetic field is perpendicular to the
principal anisotropy axis there exists a point of intersection (or degeneracy)
of the two spectrum branches with different types of spin system oscillations.
The splitting occurs due to an insignificant magnetic deflection from the basal
plane. In such a case there appears an interaction between the two types of
oscillation, the degeneracy is splitted and the magnetic field curves of resonance
frequency change in their shape.

Because of the degeneracy splitting we can observe a high-frequency branch in
polarization ({$\bf H\perp\bf h$}) and obtain the experimental dependences shown in
Fig. 5 {$\nu_2,\nu_3$}). It should be mentioned that they are little different 
in the collinear and incommensurate phases.

To study magnetic anisotropy, the angular dependences of the AFMR spectrum were
taken in planes \textbf{\textit{ac}} and \textbf{\textit{ab}}. The angular dependence of resonance
field of a low-frequency mode measured at {$\nu=78.4$ GHz} in the \textbf{\textit{ac}} plane
(rotation from axis \textbf{\textit{a}} to axis \textbf{\textit{c}} is shown in Fig. 6. As is easy to
see, there is a strong angular dependence of resonance magnetic field. Therefore,
when the angle of permanent field deflection from axis \textbf{\textit{a}} is more than 50$^\circ$,
the resonance is not observed because our experimental unit can not reach the required
resonance field.

\begin{figure}[t]
\begin{center}
\includegraphics[width=90mm]{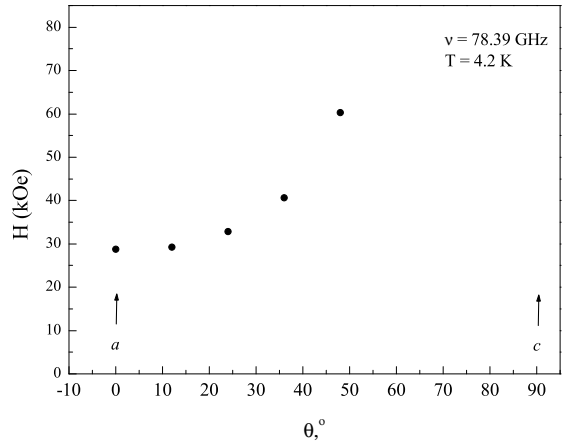}
\end{center}
\caption{The angular dependence of resonance field in the {$\textbf{\textit{ac}}$ plane} of
NdFe$_3$(BO$_3$)$_4$. T=4.2~K, $\nu$=78.4 GHz.}
\end{figure}

The measurements of angular dependences of the spectrum in the basal plane at
{$\nu=~110.75$ GHz} demonstrate a slight change in the resonance line positions along
directions {\textbf{\textit{a}} and {\textbf{\textit{b}}, suggesting the existence of magnetic anisotropy in the
plane.

To elucidate this point, the experimental dependences {$\nu_3$}(H) were taken at
frequencies below the high-frequency gap and the frequency-field dependence of
the quasi-ferromagnetic mode displayed a low-frequency gap (see Fig.5,
$\nu_3$). The absorption spectrum is shown in Fig. 7 (a low-field line).
The experimental value of gap in zero magnetic field is $\Delta\nu_{low} /\gamma=~\sqrt {2H_eH_a} =23.8$~GHz.
Knowing the value of exchange field (H$_e$=580 kOe \cite{lib5}), it is easy to estimate
the effective anisotropy field that is responsible for low-frequency gap in the basal
plane, H$_{a1}$=60 Oe. It should be noted that our value of anisotropy is much higher
than that (H$_{a1}$=12 Oe) for the Fe subsystem calculated in \cite{lib5} from the
measurements of magnetization.

At presence of a  anisotropy in the basal plane, there  should be  a spin-reorientation    
magnetic transition of spin-flop type when the external field is {$\bf H\parallel \bf a$}.
The antiferromagnetism vector is turned over perpendicular to magnetic field and the
frequency-field dependences strongly change. The evidence for this transition is
the observation of an antiferromagnetic excitation branch {$\nu_u$} (Fig. 5)
in the frequency region 21--35 GHz at H$>$10 kOe which is suggested to be due to the
excitation of AFMR "spin-flop" mode. The spectrum of the "spin-flop" mode at T=4.2 K
is shown in Fig. 7 (a high-field line). The experimental dependence for
the "spin-flop" mode of a two-sublattice antiferromagnet at magnetic field above the
spin-flop transition can be given by the expression:

\begin{equation}
(\nu_4 / \gamma)^2=H^2-{H_{sf}}^2,\label{equation2}
\end{equation}

\begin{figure}[t]
\begin{center}
\includegraphics[width=90mm]{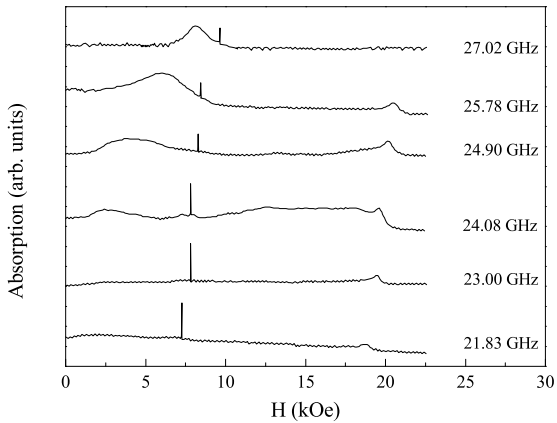}
\end{center}
\caption{The AFMR absorption spectrum at low frequencies (21--27 GHz),
T=4.2 K. The low-field line is a "quasi-ferromagnetic" branch, the
high-field one is a "spin-flop" mode. The narrow line is a DPPH signal.}
\end{figure}

As is seen from Fig. 5, the experimental field dependence at 14.6 K in the
collinear phase (open triangles) is well described by Eq. (2) while in the
incommensurate phase (4.2 K) the experimental data (dark triangles) are all out of
the theoretical dependence (2). The lack of a low-frequency oscillator made it
impossible for us to follow the frequency-field dependence of AFMR at frequencies
lower than 20 GHz.

The following peculiarity of the absorption spectrum must be emphasized: the AFMR line
width in the ordered commensurate phase ({$\Delta H$=4.3 kOe at $\bf H\parallel \bf a$ and
$\Delta H_c$=5.3 kOe at $\bf H\parallel \bf c$ for T=14.6 K}) is much larger than that
in the incommensurate system ({$\Delta H$=1.9~kOe at $\bf H\parallel \bf a$ and
$\Delta H_c$=1.54 kOe at $\bf H\parallel \bf c$ for T=12.6 K}). The spectrum is shown
in Fig. 8 and Fig. 9.

\begin{figure}[t]
\begin{center}
\includegraphics[width=90mm]{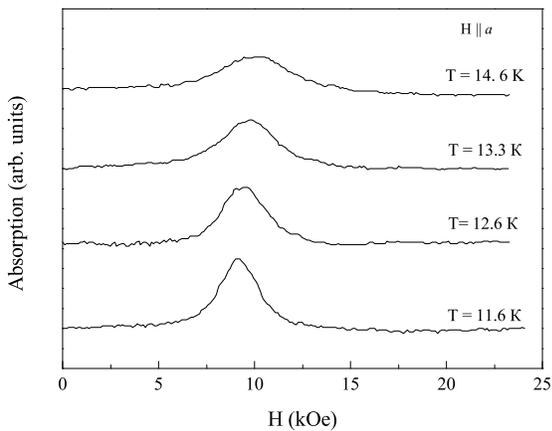}
\end{center}
\caption {The temperature dependence of AFMR line width under of
commensurate-incommensurate phase transition at {$\bf H\parallel \textbf{\textit{a}}$}.}
\end{figure}

The decrease in AFMR line width in the incommensurate phase is not much intelligible
but it may occur to the following reasons.

\begin{figure}[t]
\begin{center}
\includegraphics[width=90mm]{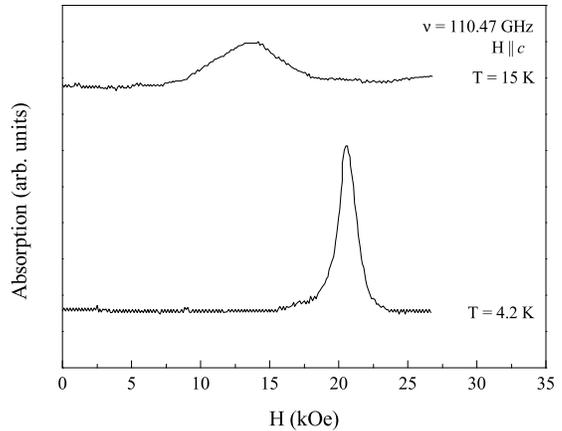}
\end{center}
\caption {Shift of resonance field and change of AFMR line width for {$\bf H\parallel \textbf{\textit{c}}$} under of commensurate-incomensurate phase transition}
\end{figure}

For the field of about 10 kOe the spin-orientation transition is followed by a number
of other effects namely, jump of magnetostriction and electrical polarization, changes
in crystal symmetry. Undoubtedly, all these effects produce modifications of relaxation
characteristics and absorption line narrowing.

It should be also emphasized that in the helicoidal modulated phase the vectors of AF
sublattice magnetization change periodically  in direction with a period of
{1140~\AA} which is incommensurate with that of the crystal lattice. This results in
that in the volume average the electrical polarization is zero unlike the homogeneous
antiferromagnetic state, besides, it may produce a AFMR line narrowing.

Thus, the analysis of the frequency-field dependences of AFMR in NdFe$_3$(BO$_3$)$_4$
at different temperatures shows that above and below the temperature of transition to
the incommensurate phase, T=13.5 K, these dependences differ considerably. An increase
in temperature up to 14.5--15 K brings the experimental and theoretical data into
coincidence for a two-sublattice collinear antiferromagnet at all orientations of
external magnetic field (see Fig. 4 and Fig. 5).

So far, we have concerned linear regimes of AFMR excitation in NdFe$_3$(BO$_3$)$_4$.
As the temperature was decreased below 4.2 K, nonlinear regimes of AFMR excitation
were revealed, i.e. nonlinear-in-absorbed power effects. The experimental conditions
were the same as those in the observation of the homogeneous linear AFMR, namely, the
external magnetic field was oriented as {$\bf H\parallel \bf a$}, the resonance excitation
frequency was 111.13 GHz and at the site of sample location in the resonator dominant
was a perpendicular polarization ({$\bf H\perp\bf h$}). The source maximum power was
5 mw. The external magnetic field was scanned in the region of absorption line, and the pump
power was varied from $-2$ to $-20$~db. The resonance curves for NdFe$_3$(BO$_3$)$_4$
at T=2~K and different levels of variable magnetic field power are shown in
Fig.10. As is seen from the figure (the two lower records), at a low
microwave radiation (lower than $-15$~db) one can observe a typical linear AFMR for the
"easy plane" --- the first resonance line corresponds to a high-frequency oscillation
mode and the second one to a low-frequency (quasi-ferromagnetic) mode. As the pumping
power is increased from $-5$ db up to $-2$ db, the quasi-ferromagnetic mode line is
broadened and one can observe a sharp jump of the resonance line amplitude. With
up~-and downward changing the direction of field there occurs a hysteresis
{$\Delta H_{hyst} \sim$ 520~Oe}, and the absorption curve displays sharp peaks.

\begin{figure}[t]
\begin{center}
\includegraphics[width=90mm]{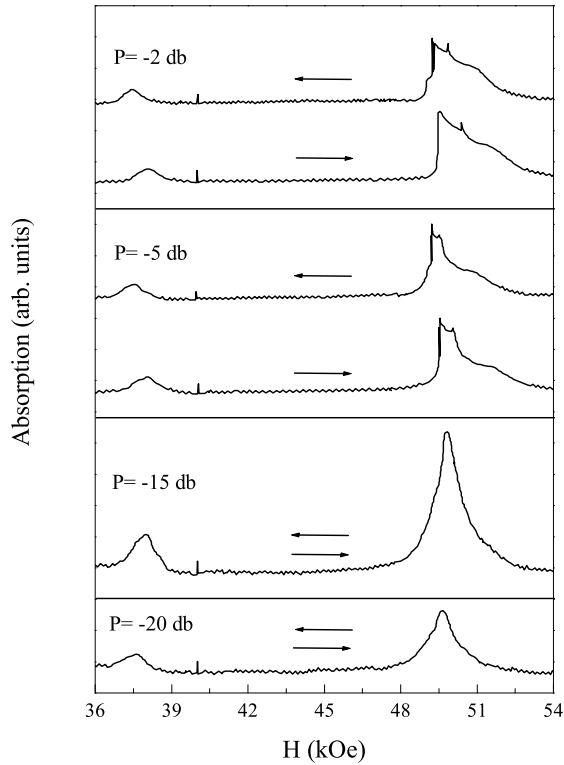}
\end{center}
\caption{The resonance curves of AFMR spectrum at different levels of
microwave power. {$\bf H\parallel \bf a$}. $\nu$=111.13 GHz, T=2.8 K.
The narrow line is a DPPH signal. The directions of field change are
shown by arrows. The spectra are shifted vertically.}
\end{figure}

It should be noted that many of the magnets studied and described in literature exhibit
nonlinear properties only at rather high pumping power (watts) \cite{lib16,lib17}, while
in NdFe$_3$(BO$_3$)$_4$ this effect is observed at a power lower than 5 mw.

\begin{figure}[t]
\begin{center}
\includegraphics[width=70mm]{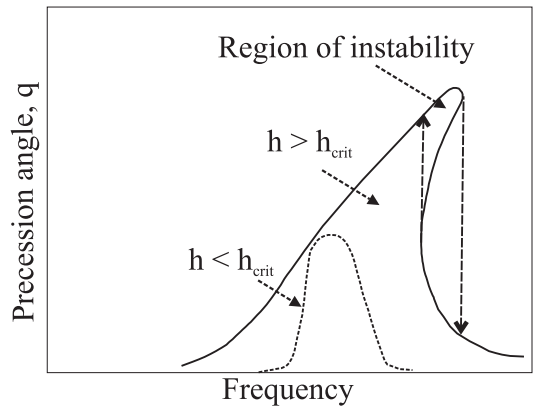}
\end{center}
\caption{The angle of magnetization precession as a function of frequency
for different levels of pumping by high-frequency variable field {$\bf h$}  \cite{lib18}.}
\end{figure}

In the theory of nonlinear resonance this phenomenon is well known; it is caused by
the dependence of nonlinear oscillation frequency on oscillation amplitude. The
essence of the effect is that at the pump field {$\bf h$} higher than some threshold
value the resonance line becomes asymmetric and ambiguous. Its top is tilted
towards lower or higher magnetic fields depending on the character of magnon interaction.
In real experiment, when static magnetic field is varied an instability occurs and a discontinuity or a jump
of the resonance line amplitude take place  \cite{lib5} \cite{lib18} (see Fig. 11). Nonlinear oscillations
appear in magnets that feature extremely low relaxation of spin excitations. At low
temperatures the nonlinear properties are observed both at ({$\bf H\parallel\bf h$}) and
({$\bf H\perp\bf h$}) polarizations when phonons are frozen out and their interaction
with magnons is highly weakened.

\section{Conclusions}
1. The magnetic field dependence of resonance frequency of AFMR spectrum of
NdFe$_3$(BO$_3$)$_4$ has been studied extensively in wide ranges of frequency
(17--142~GHz) and temperature (4.2--17.0 K). It is found that in the spectrum
of spin waves there are two branches of AFMR oscillations with considerably different
values of energy gap ({$\Delta\nu_{high}$=101.9 GHz, $\Delta\nu_{low}$=23.8 GHz})
and their associated magnetic anisotropies (H$_{a1}$=1.14 kOe, H$_{a2}$=60 Oe).

The existence of anisotropy in the basal plane results in a "spin-flop"
spin-orientation phase transition which manifests itself in the frequency-field
dependence at H $\approx10$ kOe. These experimental data suggest that according to
the type of magnetic anisotropy the compound NdFe$_3$(BO$_3$)$_4$ is an "easy-plane"
antiferromagnet with a weak anisotropy in the basal plane.

2. It has been shown experimentally that above 13.5~
K the AFMR spectrum of
NdFe$_3$(BO$_3$)$_4$ is well described in terms of the model of two-sublattice
collinear antiferromagnet.

The phase transition to an incommensurate helicoidal modulated phase at 13.5 K
is followed by an essential rearrangement of the AFMR spectrum which cannot be
described by the above-mentioned model below 13.5 K. Changes of both resonance
fields and absorption line widths are observed for all orientations of external
magnetic field.

3. Our experiments have demonstrated that the compound NdFe$_3$(BO$_3$)$_4$ is
a magnet with abnormally low spin-lattice relaxation at temperatures below 4 K
with the resulting excitation of nonlinear AFMR.

The effects of resonance instability with increasing the power of high-frequency
variable magnetic field {$\bf h$} are observed. The typical peculiarity of
NdFe$_3$(BO$_3$)$_4$ is an abnormally low threshold of nonlinear regime excitation.

  \end{document}